\documentclass[12pt,a4paper]{article}

\usepackage{amsmath,amsfonts,amssymb,bm}

\usepackage{graphicx} % do wczytywania rysunkow

\usepackage[hang,normalsize,bf]{subfigure} % rysuje kilka rysunkow
\begin{document}

\begin{center}

{\Huge \bf
Charm quark-antiquark correlations \\
in photon-proton scattering \\
}

\vspace {0.6cm}

{\large M. {\L}uszczak $^{2}$ and A. Szczurek $^{1,2}$}

\vspace {0.2cm}

$^{1}$ {\em Institute of Nuclear Physics\\
PL-31-342 Cracow, Poland\\}
$^{2}$ {\em University of Rzesz\'ow\\
PL-35-959 Rzesz\'ow, Poland\\}

\end{center}

\begin{abstract}
Correlation of charm quark - charm antiquark in $\gamma p$ scattering
are calculated in the $k_t$-factorization approach.
We apply different unintegrated gluon distributions (uGDF)
used in the literature. The results
of our calculations are compared with very recent experimental
results from the FOCUS collaboration. The CCFM uGDF 
developed recently by Kwieci\'nski et al. gives a good description
of the data.
New observables are suggested for future studies.
Predictions and perspectives for the HERA energies are presented.
\end{abstract}

In recent years a lot of activity was devoted to the description of
the photon-proton total cross section (or $F_2$ structure function)
in terms of the unintegrated gluon distribution functions (uGDF)
(see e.g.\cite{small-x-02,small-x-03} and references therein).
In some of the analyses also inclusive charm quark (or meson)
were considered \cite{QQbar}. Although the formalism of uGDF is well
suited for stydying more exclusive observables,
only very few selected cases were considered in the literature.
A special example is azimuthal jet-jet correlations
in photon-proton scattering \cite{FR94,AGKM94,SNSS}.
In this case one samples in a nontrivial way simultaneously the $x$
and $k_t$ dependences of uGDF.
The H1 collaboration at HERA has measured very recently such
correlations \cite{H1_dijets_correl}.
Very similar analysis was performed in the past also for
open charm production at somewhat lower energies \cite{correl_old}.
Recently the FOCUS collaboration at Fermilab provided new precise
data for charm-anticharm correlations \cite{FOCUS}.
It is our aim here to analyze the charm-anticharm correlations
in terms of uGDF. In the present paper we wish to compare results for
different uGDF available in the literature.
While the total cross section depends on small values of x,
the high-$p_t$ jets and/or heavy quark production test
gluon distributions at somewhat larger x.
Only some approches from the literature are applicable in this region.
In particular, we wish to test results based on
CCFM unintegrated parton distributions
developed recently by Kwieci\'nski et al.\cite{CCFM_b1,CCFM_b2}.

The total cross section for quark-antiquark production
in the reaction $\gamma + p \to Q + \bar Q + X$ can be written as
\cite{SNSS}
\begin{equation}
\sigma^{\gamma p \to Q \bar Q}(W) =
\int d \phi \int dp_{1,t}^2 \int dp_{2,t}^2 \int dz \;
\frac{f_g(x_g,\kappa^2)}{\kappa^4} \cdot
\tilde{\sigma}(W,\vec{p}_{1,t},\vec{p}_{2,t},z) \; .
\label{master_formula}
\end{equation}
In the formula above $f_g(x,\kappa^2)$ is the unintegrated gluon
distribution with the convention from Ref.\cite{SNSS}.
\footnote{Two different conventions are used throughout the literature
\cite{small-x-02}.}
The gluon transverse momentum is related to the quark/antiquark
transverse momenta $\vec{p}_{1,t}$ and $\vec{p}_{2,t}$ as:
\begin{equation}
\kappa^2 = p_{1,t}^2 + p_{2,t}^2 + 2 p_{1,t} p_{2,t} cos\phi \; .
\end{equation}
In Eq.(\ref{master_formula}) we have introduced:
\begin{eqnarray}
\tilde{\sigma}(W,\vec{p}_{1,t},\vec{p}_{2,t},z)
= \frac{\alpha_{em}}{2} \; e_Q^2 \; \alpha_s(\mu_r^2) \nonumber \\
\left\{
   [ z^2+(1-z)^2 ] \;
  \Bigg\vert \; \frac{\vec{p}_{1,t}}{p_{1,t}^2 + m_Q^2} +
     \frac{\vec{p}_{2,t}}{p_{2,t}^2 + m_Q^2} \;
  \Bigg\vert ^2
   \; + \;
   m_Q^2 \left(
   \frac{1}{p_{1,t}^2 + m_Q^2} +
   \frac{1}{p_{2,t}^2 + m_Q^2} +
      \right)^2 
\right\}   \; .
\label{aux}
\end{eqnarray}
The unintegrated gluon distribution $f_g$ is evaluted at
\begin{equation}
x_g = \frac{M_t^2}{W^2} \; ,
\label{x_g}
\end{equation}
where
\begin{equation}
M_t^2 = \frac{p_{1,t}^2 + m_Q^2}{z} + \frac{p_{2,t}^2 + m_Q^2}{1-z}
\; .
\label{invariant_mass}
\end{equation}
It is obvious that at larger transverse momenta of quarks and/or
heavy quark-antiquark production one samples larger values of $x_g$
than in the case of the total photon-proton cross section.

The choice of $\mu_r^2$ is not essential
for the discussion in the present paper and will be discussed
elsewhere.
In the present calculation the scale of running coupling constant
in (\ref{aux}) is taken to be $\kappa^2$ and the freezing prescription
from \cite{SS} is used. The latter prescriptions are not
very important when studying correlations. They may
be important, however, for the integrated cross sections,
to be studied elsewhere \cite{LS04}.

Thus the basic ingredient of our approach are unintegrated gluon
distributions. Different models of uGDF have been proposed
in the literature (see for instance \cite{small-x-02,small-x-03}
and references therein). The main effort has been concentrated
on the small-x region. While the total cross section is the genuine
small-x phenomenon (x $<$ 10$^{-3}$), the production of
charm and bottom quarks samples rather the intermediate-x region
(x$\sim$ 10$^{-2}$ - 10$^{-1}$) even at the largest available
energies at HERA. It is not obvious a priori
if the methods used are appropriate for the intermediate values of x.
In the present approach we shall present results for a few
selected gluon distributions from the literature.
For illustration we shall consider the simple BFKL \cite{BFKL},
the saturation model used to study HERA photon-proton total cross
sections \cite{GBW_glue} (GBW),
and the saturation model being often used recently to calculate
particle production in hadron-hadron collisions \cite{KL01} (KL).
These three model approaches are expected to be valid for small,
not very well specified, values of x.
At somewhat larger values of x all these models
are expected to break. This may happen already at the FOCUS energy
W = 18.4 GeV. Clearly an extrapolation may be needed.
As in Ref.\cite{pp_pions} one can try to extend
the applicability of these small-x models by multiplying the model
distributions by a phenomenological factor $(1-x)^n$.
In principle, the value of $n$ could be adjusted to inclusive
spectra at lower energies. The choice of $n$ is, however,
marginal for the correlations studied in the present paper. 
We shall not discuss here the details of the different approaches.
A more detailed discussion can be found in Ref.\cite{pp_pions}.

In addition, we shall consider two other approaches adequate
for intermediate-x region. The CCFM approach seems to be
the best tailored for this purpose.
It was shown in Refs.\cite{CCFM_b1,CCFM_b2} how to solve the
one-loop CCFM equation in the impact parameter representation.

The unintegrated parton distributions used in the present
paper were obtained by solving the Kwieci\'nski CCFM equations
\cite{CCFM_b1,CCFM_b2} using LO GRV98 collinear distributions
\cite{GRV98} as the input for the evolution.
By construction this procedure assures that our uPDF provide
a good description of the $F_2$ structure function data.

The solution of the CCFM equation depends on three variables
${\tilde f}_g = \tilde f_g(x,b,\mu^2)$.
\footnote{In the present paper we shall use the notation $\tilde f$
instead of $\bar f$ as in Refs.\cite{GKB03,GK03}.}
The familiar momentum representation unintegrated gluon distribution can
be obtained via Fourier-Bessel transform
\begin{equation}
f_q(x,\kappa^2,\mu^2) = \frac{1}{2 \pi}
 \int  \exp \left( i \vec{\kappa} \vec{b} \right) \; 
\tilde f_q(x,b,\mu^2) \; d^2 b \; .
\label{Fourier transform}
\end{equation}

As already mentioned in the introduction, it is our intention here
to use uGDFs $\tilde{f}_g^{CCFM}(x,b,\mu^2)$ which fulfill the b-space
(one-loop) CCFM equations \cite{CCFM_b1,CCFM_b2}.
However, the perturbative solution $\tilde{f}_g^{CCFM}(x,b,\mu^2)$
does not include nonperturbative effects such as, for instance,
intrinsic momentum distribution of partons in colliding hadrons.
In order to include such effects we propose to modify the perturbative
solution $\tilde{f}_g^{CCFM}(x,b,\mu^2)$
and write the modified gluon distribution
$\tilde{f}_g(x,b,\mu^2)$ in the simple factorized form
\begin{equation}
\tilde{f}_g(x,b,\mu^2) = \tilde{f}_g^{CCFM}(x,b,\mu^2)
 \cdot F_g^{NP}(b) \; .
\label{modified_uPDFs}
\end{equation}
In Ref.\cite{KS04}, two
different functional forms for the nonperturbative form factor
\begin{equation}
F_g^{NP}(b) = F^{NP}(b) = \exp\left(- \frac{b^2}{4 b_0^2}\right) \;
\text{or} \; \exp \left( - \frac{b}{b_e} \right)
\label{formfactor}
\end{equation}
identical for all species of partons were used.
In Eq.(\ref{formfactor}) $b_0$ (or $b_e$) is the only free parameter.
The parameters were roughly adjusted in \cite{KS04} to describe
production of W and Z bosons in nucleon-nucleon collisions.
In the present note we shall show only results for the Gaussian
form factor. The dependence on the choice of the form factor
will be studied elsewhere.

The resummation formulae \cite{CSS85} and the unintegrated parton
distribution formulae for Higgs \cite{GK03} and gauge boson \cite{KS04}
have identical structure if the following formal assignment is made:

\begin{eqnarray}
{\tilde f}_{g}^{SGR}(x,b,\mu^2) &=& \frac{1}{2}
 F_g^{NP}(\mu,b,x) 
\left[ x g(x_1,\mu(b)) + ... \right]
\exp \left( \frac{1}{2} S_g\left( b,\mu \right) \right) \; , \nonumber \\
{\tilde f}_{q}^{SGR}(x,b,\mu^2) &=& \frac{1}{2}
 F_q^{NP}(\mu,b,x) 
\left[ x q(x_1,\mu(b)) + ... \right]
\exp \left( \frac{1}{2} S_q\left( b,\mu \right) \right) \; , \nonumber \\
{\tilde f}_{\bar q'_2}^{SGR}(x,b,\mu^2) &=& \frac{1}{2}
 F_{\bar q'}^{NP}(\mu,b,x)
\left[ x \bar q(x,\mu(b)) + ... \right]
\exp \left( \frac{1}{2} S_{\bar q'} \left( b,\mu \right) \right)  \; .
\label{effective_uPDF}
\end{eqnarray}
The index $SGR$ above stands for ``soft-gluon resummation''.
The explicit expressions for $S_g$, $S_q$ and $S_{\bar q}$ can be
found in Ref.\cite{GK03,KS04}, where in addition similarities and
differences between Kwieci\'nski CCFM and soft gluon resummation
are discussed.

The $k_t$-dependent unintegrated distributions of gluons corresponding
to the b-space resummation can be then obtained through
the Fourier-Bessel transform
\begin{equation}
f_g^{SGR}(x,\kappa^2,Q^2) = \int db b \; J_0(\kappa b)
\tilde{f}_g^{SGR}(x,b,Q^2).
\label{standard_uPDF_SGR}
\end{equation}

With the simple Ansatz (\ref{formfactor}) for $F_g^{NP}$
the whole scale $\mu^2$ dependence resides exclusively in
the Sudakov-like form factor.
For brevity, we shall call the gluon distribution in
Eq.(\ref{standard_uPDF_SGR}) the ``resummation gluon distribution''.

\begin{figure}[htb] % Figure 1
  \subfigure[]{\label{fig_pt2_low}
    \includegraphics[width=7.0cm]{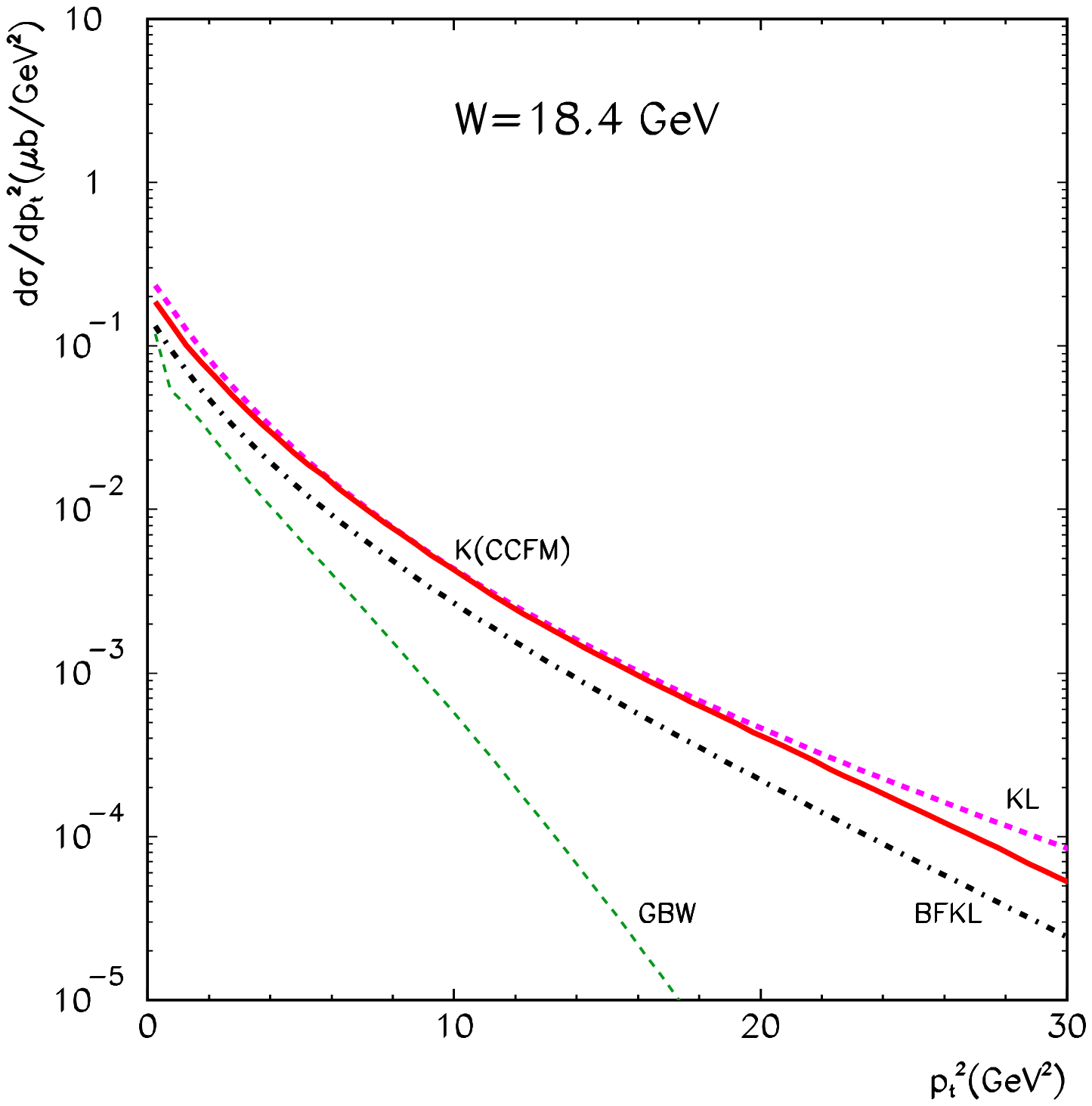}}
  \subfigure[]{\label{fig_pt2_high}
    \includegraphics[width=7.0cm]{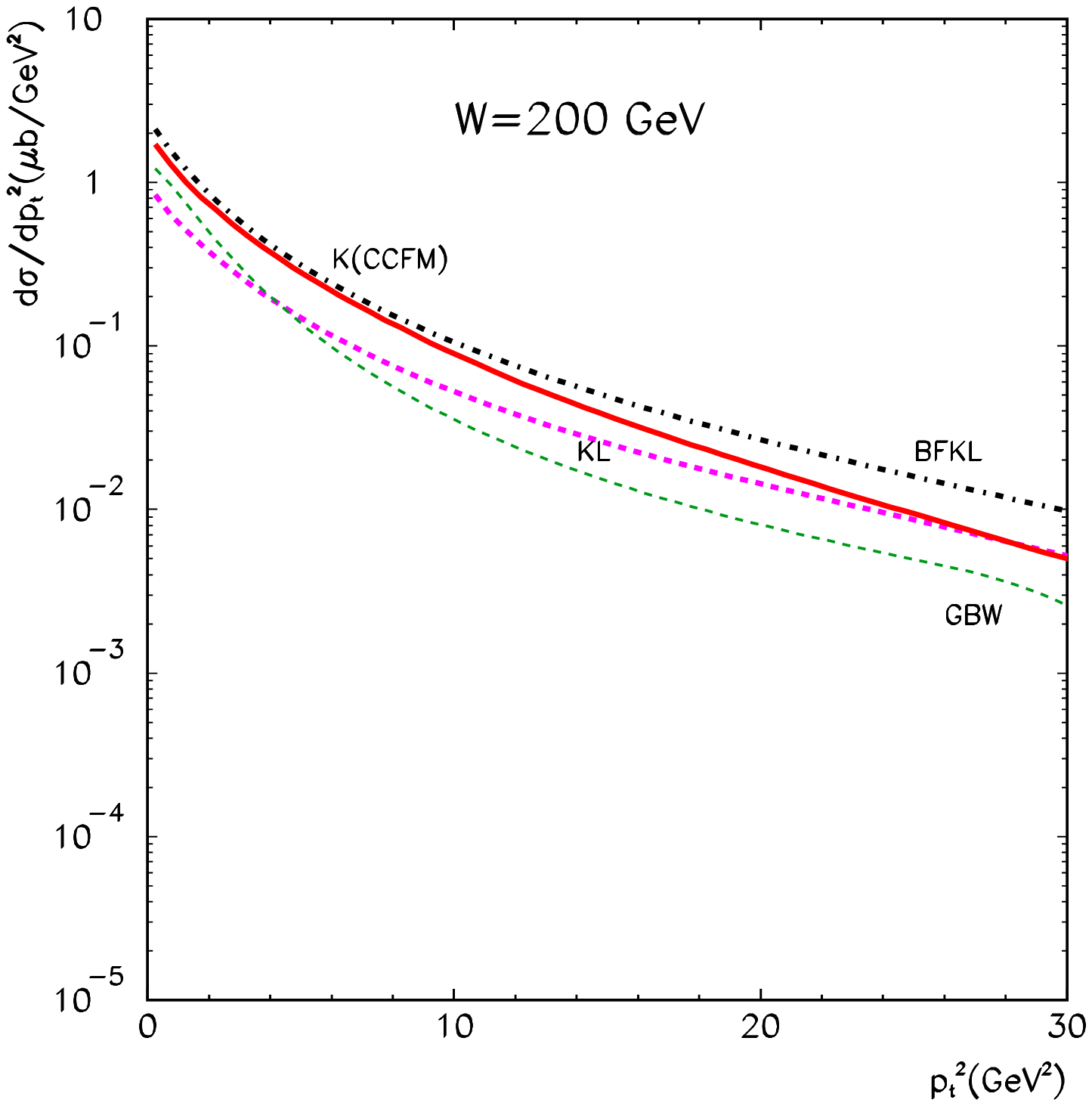}}
%\begin{center}
%\includegraphics[width=8cm]{fig1.eps}
\caption{\it
Inclusive distribution of charm quarks/antiquarks as a function
of the corresponding transverse momentum.
We present results for GBW (thin dashed), KL (thick dashed),
BFKL (dash-dotted) and Kwieci\'nski CCFM (solid).
\label{fig_pt2}
}
%\end{center}
\end{figure}

Before we go to charm quark - charm antiquark correlations
we wish to show the results for inclusive spectra of $c$ or $\bar c$.
In Fig.\ref{fig_pt2} we show the distributions
$\frac{d \sigma}{d p_{t}^2}$ for different uGDF for low
(W=18.4 GeV, left panel) and high (W=200 GeV, right panel) energies.
At the lower energy the GBW uGDF gives somewhat steeper $p_t^2$
distribution than the other uGDF.
At the high energy the slope of the $p_t^2$ distributions decreases.
Quite similar slopes are obtained for different uGDF.
Therefore this observable is not the best one to test models of uGDF.
We shall show below that more exclusive correlation observables
are more sensitive tests of models/parametrizations of uGDF.

In the present analysis we do not put any restrictions on heavy quark
or heavy antiquark transverse momenta
$\vec{p}_{1,t}$ and $\vec{p}_{2,t}$. A detailed analysis of
the effect of such cuts on the results will be presented
elsewhere \cite{LS04}.

\begin{figure}[htb] % Figure 2
  \subfigure[]{\label{fig_phi_low}
    \includegraphics[width=7.0cm]{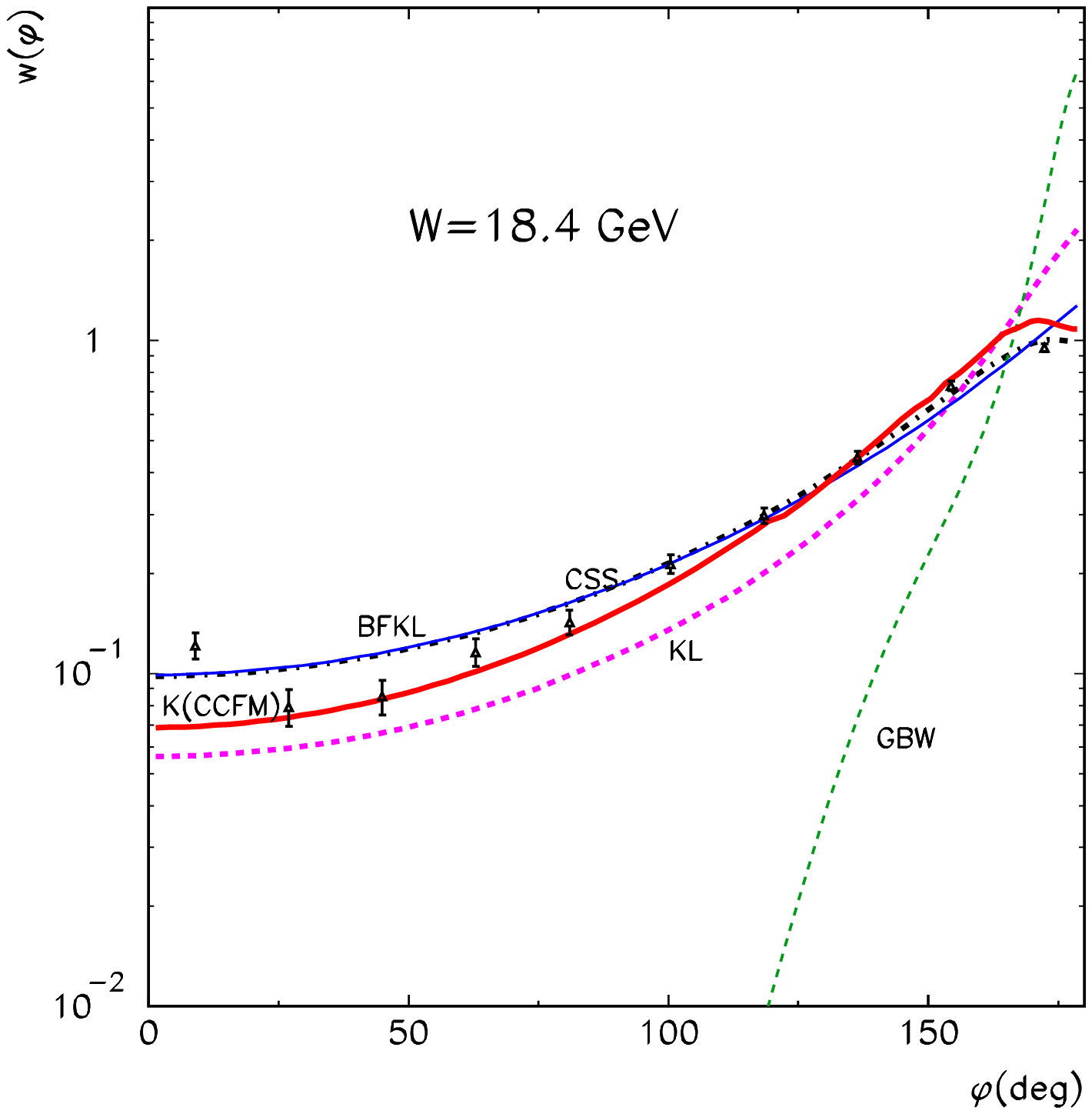}}
  \subfigure[]{\label{fig_phi_high}
    \includegraphics[width=7.0cm]{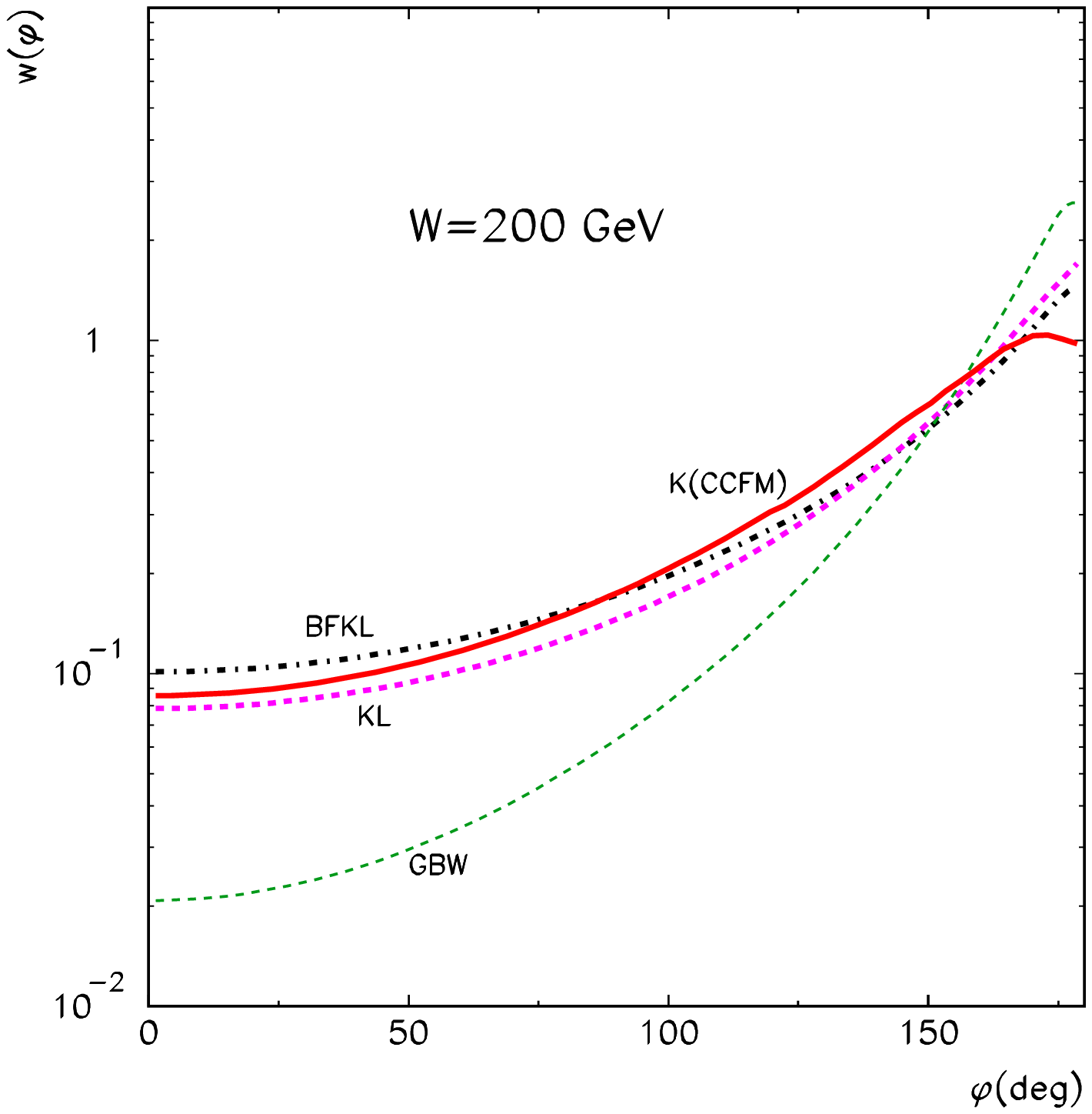}}
%\begin{center}
%\includegraphics[width=8cm]{fig1.eps}
\caption{\it
Azimuthal correlations between $c$ and $\bar c$.
The theoretical results are compared to the recent results
from \cite{FOCUS} (fully reconstructed pairs).
The notation is the same as in Fig.\ref{fig_pt2}.
\label{fig_phi}
}
%\end{center}
\end{figure}

The azimuthal correlation functions $w(\phi)$
defined as:
\begin{equation}
w(\phi;W) \equiv \frac{\frac{d\sigma}{d \phi}(W)}
{\int \frac{d \sigma}{d \phi}(W)\; d \phi} \; ,
\end{equation}
where
\begin{equation}
\frac{d \sigma}{d \phi}(W) =
 \int dp_{1,t}^2 \int dp_{2,t}^2 \int dz \;
\frac{f_g(x_g,\kappa^2)}{\kappa^4} \cdot
\tilde{\sigma}(W,\vec{p}_{1,t},\vec{p}_{2,t},z)
\end{equation}
in order that
\begin{equation}
\int w(\phi;W) \; d \phi = 1
\label{w_renor}
\end{equation}
for two energies of $W$ = 18.4 GeV (FOCUS)
and $W$ = 200 GeV (HERA) are shown in Fig.\ref{fig_phi}.
The GBW-glue (thin dashed) gives too strong back-to-back correlations
for the lower energy. Another saturation model (KL, \cite{KL01})
provides more angular decorrelation, in better agreement with
the experimental data. The BFKL-glue (dash-dotted) provides very
good description of the data. The same is true for the
CCFM-glue (thick solid) and resummation-glue (thin solid).
The latter two models are more adequate for the lower energy.
The renormalized azimuthal correlation function (\ref{w_renor})
for BFKL, GBW and KL models are almost independent of the power
$n$ in extrapolating to larger values of $x_g$.
In the present paper, in calculating the cross section with
the Kwieci\'nski CCFM uGDF for simplicity
we have fixed the scale for $\mu^2$ = 4 $m_c^2$.
Allowing for dependence of the scale $\mu^2$ on kinematical variables
such as $x$ or $\kappa^2$ would make the calculation very time
consuming. The sensitivity to the choice of the scale will
be discussed in detail elsewhere.
For comparison in panel (b) we present predictions for $W$ = 200 GeV.
Except of the GBW model, there is only a small increase of
decorrelation when going from the lower fixed-target energy region
to the higher collider-energy region.

In the present paper we completely ignore the resolved photon
component \cite{szczurek02}. The latter should be, however,
negligible for the FOCUS fixed target experiment \cite{FOCUS}.

\begin{figure}[htb] % Figure 3
  \subfigure[]{\label{fig_map_gbw}
    \includegraphics[width=7.0cm]{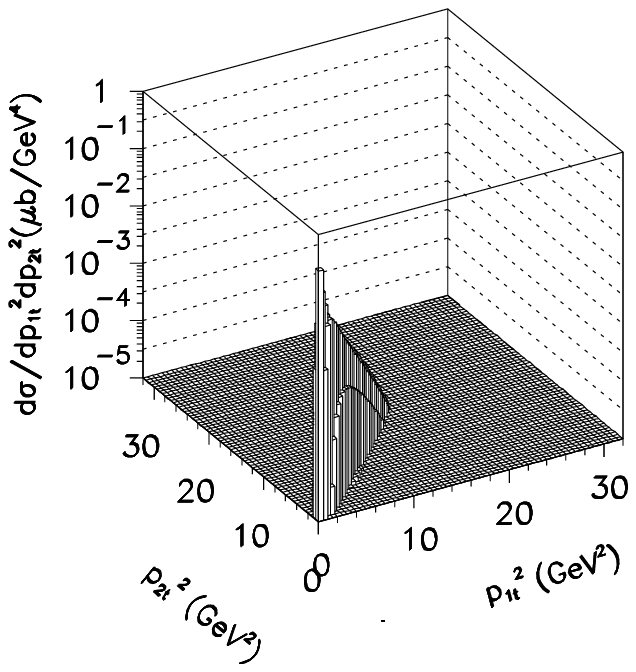}}
  \subfigure[]{\label{fig_map_kl}
    \includegraphics[width=7.0cm]{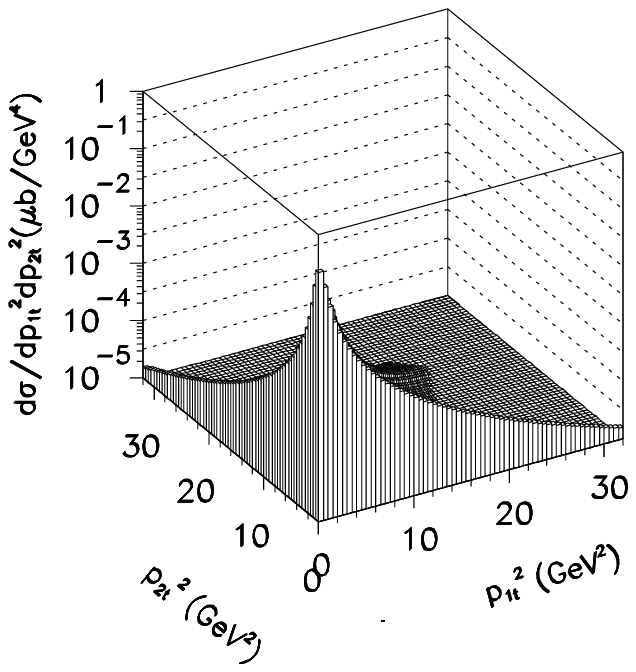}}
  \subfigure[]{\label{fig_map_bfkl}
    \includegraphics[width=7.0cm]{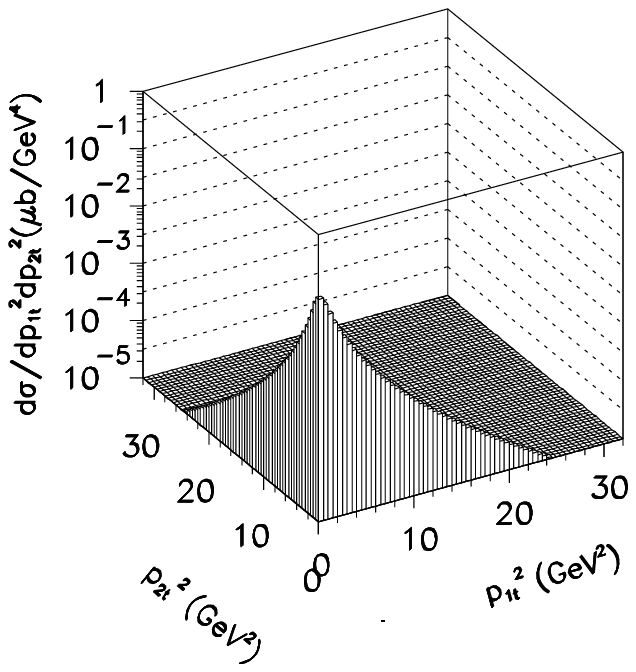}} 
  \subfigure[]{\label{fig_map_ccfm}
    \includegraphics[width=7.0cm]{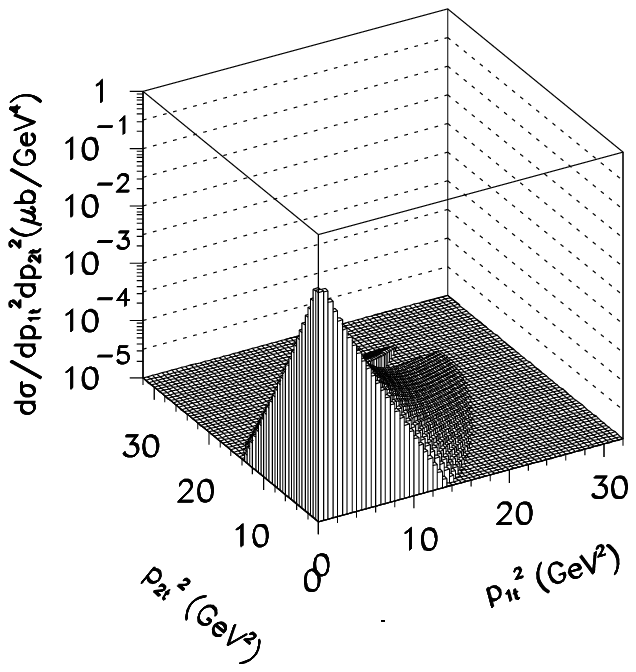}}
  \subfigure[]{\label{fig_map_ccs}
    \includegraphics[width=7.0cm]{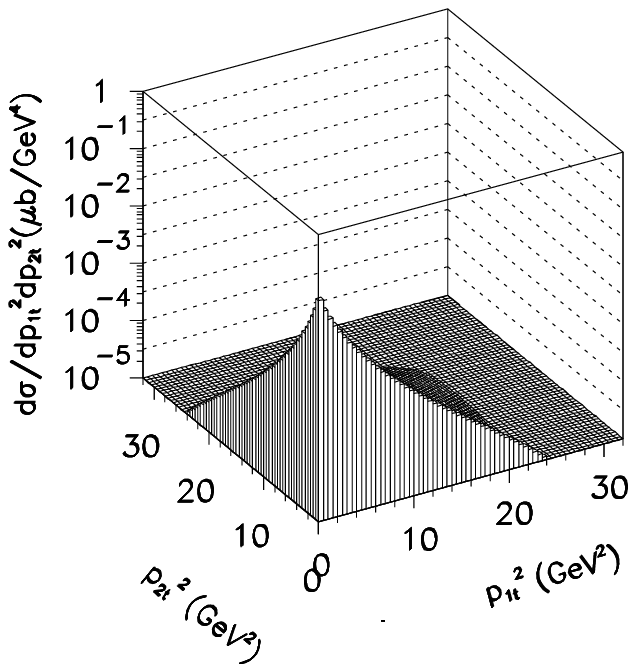}}
\caption{\it
Some examples of the two-dimensional maps $w(p_{1,t}^2,p_{2,t}^2)$
for W = 18.4 GeV: a) GBW, b) KL, c) BFKL, d) CCFM, e) resummation.
\label{fig_maps}
}
\end{figure}

Not only azimuthal correlations are interesting.
In general, the integrand of Eq.(\ref{master_formula}) depends
on four independent kinematical variables
$\phi, p_{1,t}^2, p_{2,t}^2, z$ (other combinations of the kinematical
variables are also possible). In particular, the formula
(\ref{master_formula}) can be rewritten in the form
\begin{equation}
\sigma^{\gamma p \to Q \bar Q}(W) =
\int dp_{1,t}^2 \int dp_{2,t}^2 \;
w(p_{1,t}^2,p_{2,t}^2;W)
\label{p1t2_p2t2_correlations}
\end{equation}
where the two-dimensional correlation function
\begin{equation}
w(p_{1,t}^2,p_{2,t}^2;W) =
\int d \phi \; dz  \;
\frac{f_g(x_g,\kappa^2)}{\kappa^4} \cdot
\tilde{\sigma}(W,\vec{p}_{1,t},\vec{p}_{2,t},z) \; .
\label{w_p1t2_p2t2}
\end{equation}
In Fig.\ref{fig_maps} we present some examples of
$w(p_{1,t}^2, p_{2,t}^2)$ at W=18.4 GeV for different uGDF.
%As in Ref.\cite{szczurek02} for GBW, KL and BFKL we have used
%the power n=7 in the extrapolating formula.
%While the normalization is slightly dependent on the value of the power,
%the shape of the two-dimensional map is almost the same.
The maps for different uGDF differ in details.
The distribution for the GBW gluon distribution is concentrated along
the diagonal $p_{1,t}^2 = p_{2,t}^2$ and in this respect resembles
the familiar collinear leading-order result.
\footnote{It is well known that the collinear NLO calculation
is not reliable for $p_{1,t} = p_{2,t}$.}
The other three distributions have a sizeable strength at the
phase-space borders for
$p_{1,t}^2 \approx$ 0 or $p_{2,t}^2 \approx$ 0.
The KL gluon distribution gives in addition some enhancement
at $p_{1,t}^2 \approx p_{2,t}^2$, especially at large
transverse momenta. Experimental 
studies of such maps could open an interesting new possibility
to test models of uGDF in a more detailed differential fashion.
In principle, such studies will be possible with HERA II runs
at DESY.
The leading-order approach of Kwieci\'nski
contains the higher-order corrections via evolution equations.
However, in contrast to NLO collinear approach it can be applied
even in the region $p_{1,t} = p_{2,t}$.

At present experimental luminosities (statistics) one
may have a problem to explore the whole two-dimensional maps
shown in Fig.\ref{fig_maps}.
In this case a more global variable could be useful.
In order to quantify the spread over $p_{1,t}^2 \times p_{2,t}^2$ plane
and/or departure from the diagonal (LO collinear approach)
we propose a new variable which can be interpreted as
a measure of deviations from the equal-length momenta
defined as:
\begin{equation}
f(p_{max}^2 > k \; p_{min}^2;W) \equiv
\frac{\sigma(p_{max}^2 > k \; p_{min}^2;W)}{\sigma(W)} \; ,
\label{f_noncollinear}
\end{equation}
where $p_{max}^2 = \max(p_{1,t}^2,p_{2,t}^2)$
and $p_{min}^2 = \min(p_{1,t}^2,p_{2,t}^2)$.
For example with $k$ = 2 one obtains respectively:
GBW: 0.01, KL: 0.45, BFKL: 0.63, K(CCFM): 0.60, resum: 0.33.
The quantity $f(p_{max}^2 > k p_{min}^2; W)$ is sensitive to
both perturbative ($p_{1,t} \ne p_{2,t}$)
and nonperturbative ($p_{1,t} \approx p_{2,t}$) processes and
therefore reflects their interplay.
We believe that the FOCUS collaboration could reprocess their
present data in order to obtain analogous fractions
for their ($p_{t,D}^2,p_{t,\bar D}^2$) distributions.

\begin{figure}[htb] % Figure 4
  \subfigure[]{\label{fig_psum2_low}
    \includegraphics[width=7.0cm]{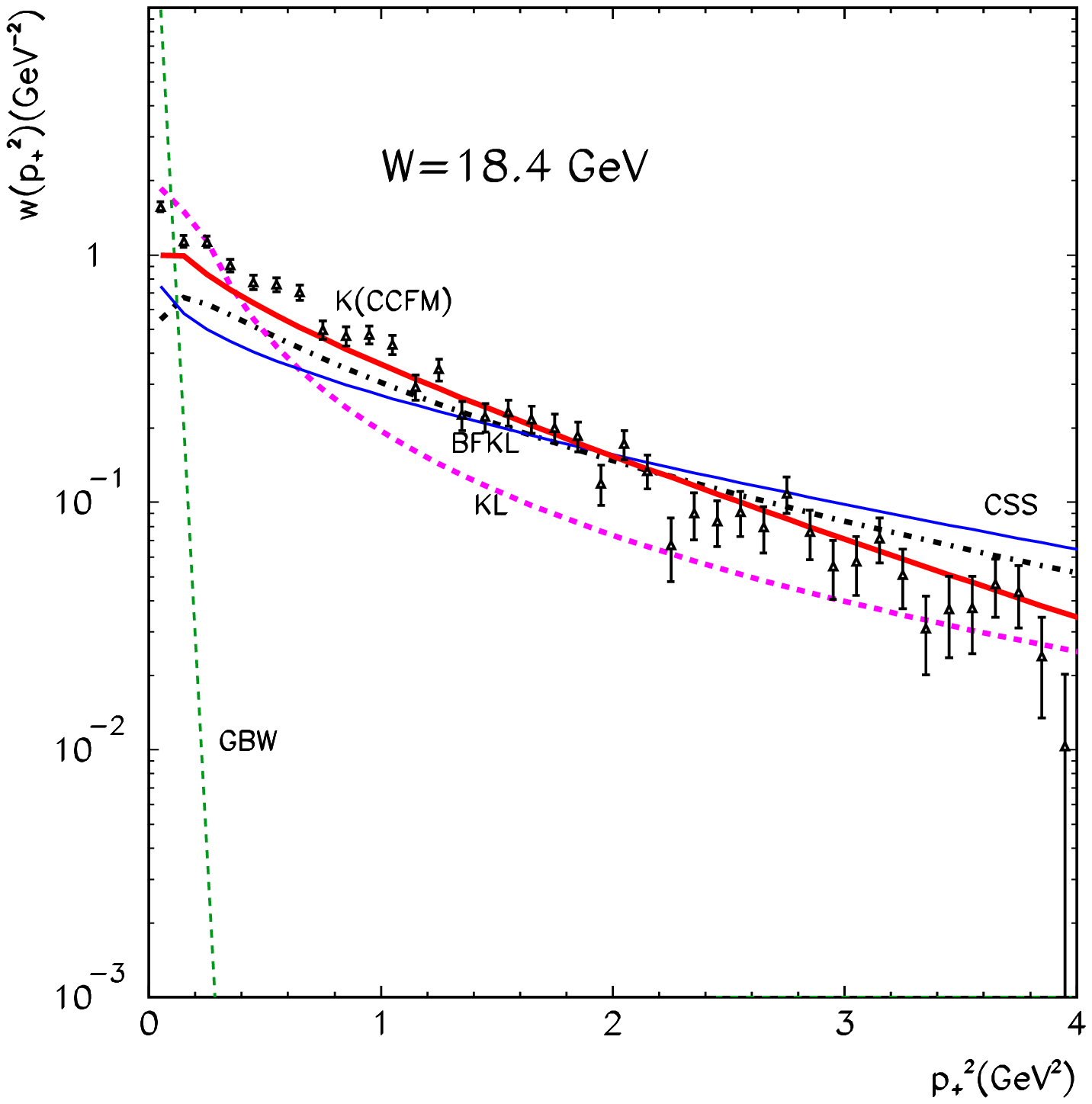}}
  \subfigure[]{\label{fig_psum2_high}
    \includegraphics[width=7.0cm]{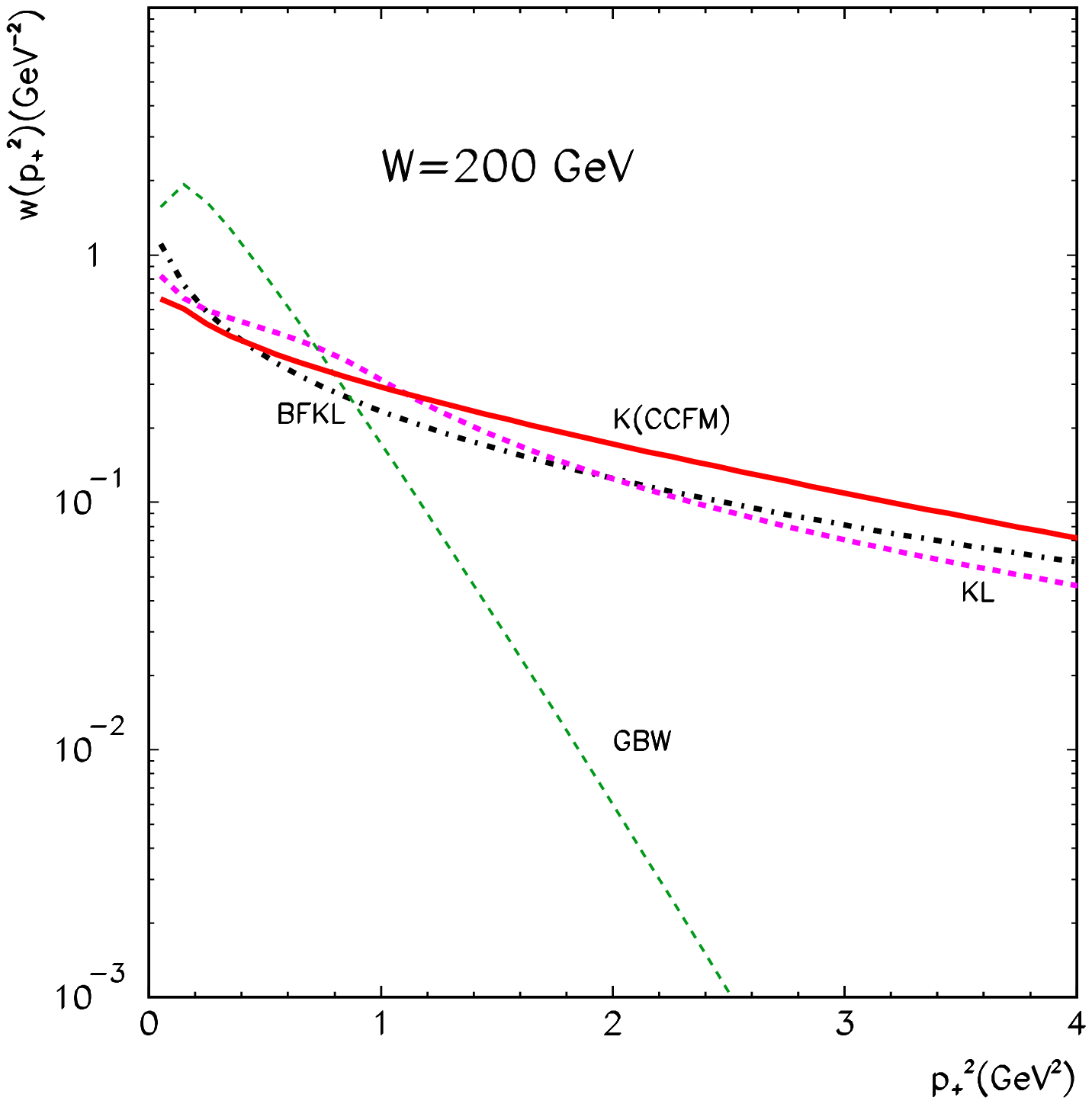}}
%\begin{center}
%\includegraphics[width=14cm]{fig2.eps}
\caption{\it
$p_+^2$ distribution of $c - \bar c$. 
The theoretical results are compared to the recent results
from \cite{FOCUS} (fully reconstruced pairs).
The notation is the same as in Fig.\ref{fig_pt2}.
\label{fig_psum2}
}
%\end{center}
\end{figure}

In the leading-order collinear approach (without parton showers
included) the transverse momenta of two jets add up to zero.
It is not the case for our leading-order
$k_t$-factorization approach. In Fig.\ref{fig_psum2} we present
normalized to unity distribution in $p_{+}^2$, where
$\vec{p}_{+} = \vec{p}_1 + \vec{p}_2$. Due to momentum conservation,
in our approach the sum of
transverse momenta is directly equal to transverse momentum of
the gluon ($\vec{p}_{+} = \vec{\kappa}$). This means that
the distribution in $p_{+}^2$ directly probes the transverse
momentum distribution of gluons.
The situation here is very similar to the situation for the azimuthal
angle distribution.
The GBW gluon distribution which discribes very well the total
$\gamma^*-p$ cross section gives very steep distribution in $p_+^2$
in comparison to other uGDF. We expect that the inclusion
of QCD evolution effects, like in Ref.\cite{BGK02} for instance,
should change this result and lead to somewhat broader distributions.
The Kwieci\'nski CCFM gluon distribution gives the best
description of the FOCUS data \cite{FOCUS}. We expect that this
approach is suitable for $x > 0.01$. Although the other models
give also reasonable description of the FOCUS $w(p_{+}^2)$ data,
one should remember that their application for the low-energy data
is somewhat unsure. 

\begin{figure}[htb] % Figure 5
  \subfigure[]{\label{fig_pdif2_low}
    \includegraphics[width=7.0cm]{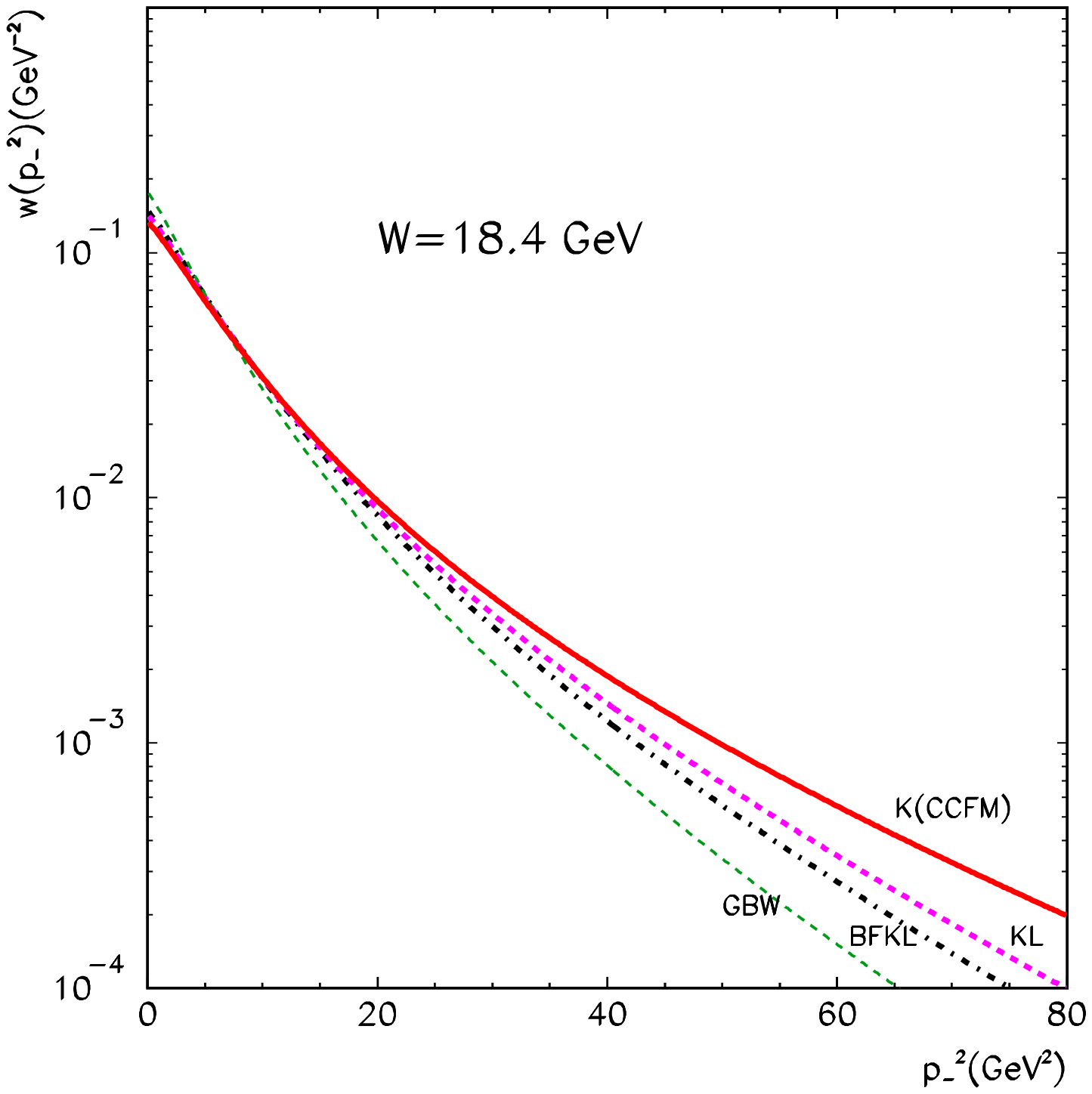}}
  \subfigure[]{\label{fig_pdif2_high}
    \includegraphics[width=7.0cm]{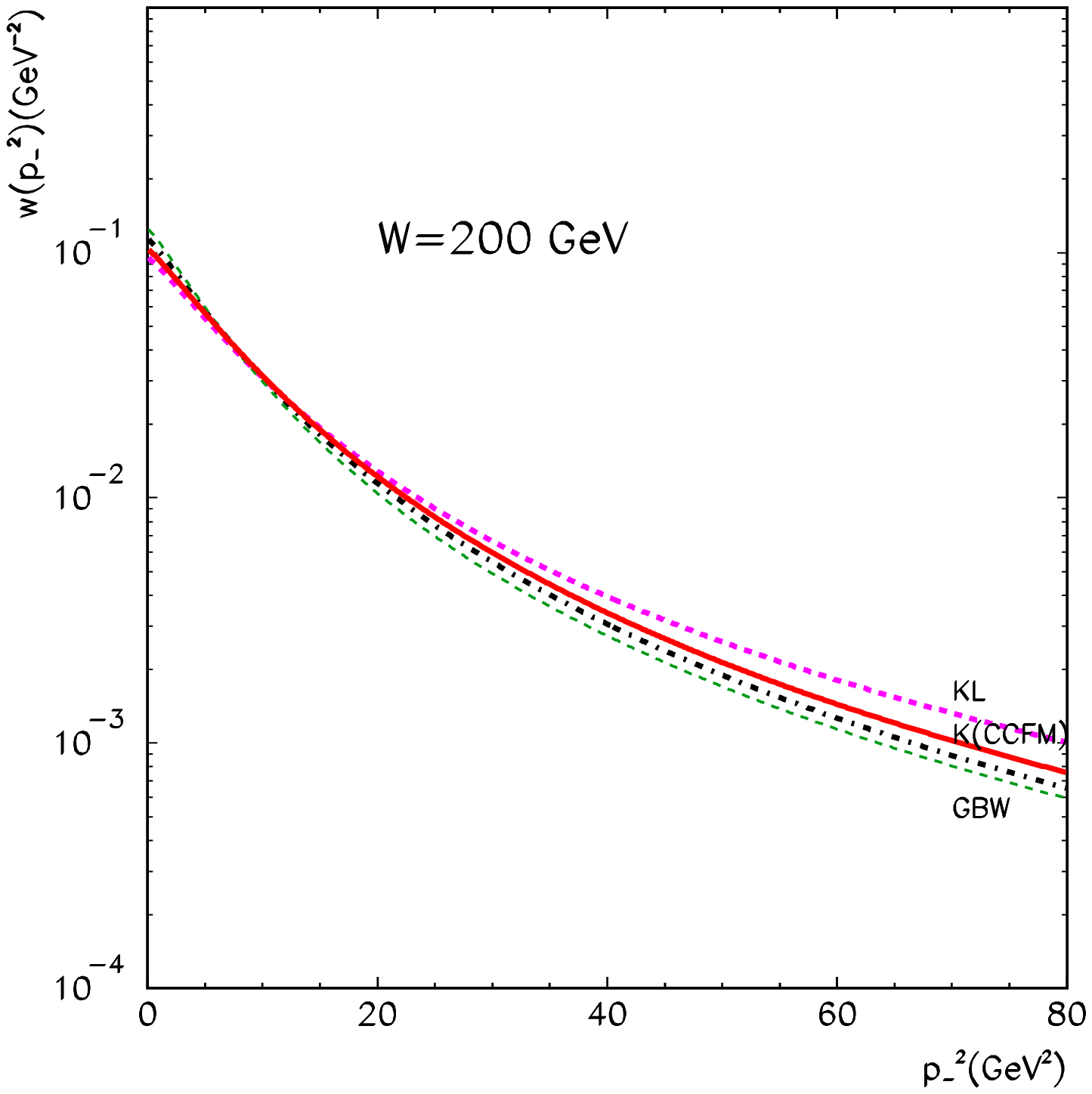}}
%\begin{center}
%\includegraphics[width=14cm]{fig3.eps}
\caption{\it
$p_{-}^2$ distribution of $c - \bar c$. 
The notation is the same as in Fig.\ref{fig_pt2}.
\label{fig_pdif2}
}
%\end{center}
\end{figure}

For completeness in Fig.\ref{fig_pdif2} we present
normalized to unity distributions in the square of
$\vec{p}_{-} = \vec{p}_{1} - \vec{p}_{2}$. Here, however,
the differences between different uGDF are smaller than
in the previous two cases. Nevertheless the corresponding data
would be a new possible observable to verify
the unintegrated gluon distributions.

In summary, we have shown that the analysis of kinematical correlations
of charm quarks and antiquarks opens new possibilities
for verifying models of uGDF.
The recently measured data of the FOCUS collaboration at Fermilab
allows one to study the unintegrated gluon distribution in the
intermediate-x region. Many models of uGDF used
in the literature are constructed rather for small values of x and
its application in the region of somewhat larger x (x $>$ 0.05)
is questionable.
The unintegrated gluon (parton) distribution which fulfill
the CCFM equations, developed recently by Kwieci\'nski et al.,
describe the data fairly well.
It can be expected that the correlation data from the HERA II runs
will give a new possibility to verify the different models of
unintegrated gluon distributions in a more detailed way.

\vskip 1cm

{\bf Acknowledgments}
We are indebted to Krzysztof Golec-Biernat and 
Hannes Jung for a discussion and Erik Gottschalk
for providing us files with the FOCUS experimental data.

%------------------------------------------------------------------

%-----------------------------------------------------------------

%------------------------------------------------------------------

%----------------------------------------------------------------

%----------------------------------------------------------------

%------------------------------------------------------------------

%------------------------------------------------------------------

\end{document}